\documentclass{article}
\usepackage{amsmath}
\usepackage{amssymb}
\usepackage{xcolor}
\usepackage[]{lineno}
\usepackage{amsmath}
\usepackage{geometry}
\usepackage{mathtools, nccmath}
\usepackage{graphicx}
\usepackage{hyperref}
\usepackage{xcolor}
\usepackage[]{lineno}
\usepackage{amsmath}
\usepackage{geometry}
\usepackage{mathtools, nccmath}
\usepackage{graphicx}
\usepackage{verbatim}
\usepackage{authblk}
\usepackage[utf8]{inputenc}
\usepackage[english]{babel}
\usepackage{natbib}
\setcitestyle{authoryear,open={(},close={)}}
\usepackage{enumitem}

\title{Use of neural networks for stable, accurate and physically consistent parameterization of subgrid atmospheric processes with good performance at reduced precision}

\author{Janni Yuval}
\author{Paul A. O'Gorman}
\author{Chris N. Hill}
\affil[1]{Massachusetts Institute of Technology, Cambridge, Massachusetts 02139, USA }

\begin{document}
\date{}
\maketitle
\date{}

\begin{abstract} 
A promising approach to improve climate-model simulations is to 
replace traditional subgrid parameterizations based on simplified 
physical models by machine learning algorithms that are data-driven.  
However, neural networks (NNs) often lead to instabilities and climate 
drift when coupled to an atmospheric model.  Here we learn an NN parameterization 
from a high-resolution atmospheric simulation in an idealized domain by coarse 
graining the model equations and output.  The NN parameterization has a 
structure that ensures physical constraints are respected, and it leads 
to stable simulations that replicate the climate of the high-resolution 
simulation with similar accuracy to a successful random-forest 
parameterization while needing far less memory.  
We find that the simulations are stable for a variety of NN 
architectures and horizontal resolutions, and that an NN with 
substantially reduced numerical precision could decrease computational 
costs without affecting the quality of simulations.

\end{abstract}

\section{Introduction}

\textcolor{black}{
Traditional parameterizations  in general circulation models (GCMs) rely on simplified physical models and suffer from inaccuracies which 
lead to model biases 
and large uncertainties in climate projections
\citep{bony2005marine,wilcox2007frequency,ogorman2012sensitivity,oueslati2015double,schneider2017climate}.
One alternative to traditional parameterizations is to use machine learning (ML) algorithms to create new subgrid parameterizations \citep{krasnopolsky2013using,gentine2018could,rasp2018deep,
brenowitz2018prognostic,Ogorman2018using,
brenowitz2019spatially,bolton2019applications,yuval2020stable,han2020moist,zanna2020data}.}

Two ML algorithms that have been used for 
climate-model parameterizations are neural networks (NNs) and random forests (RFs).
\citet{Ogorman2018using} trained an RF to emulate a conventional convection scheme of an atmospheric GCM and showed that when this RF parameterization is implemented in the GCM it leads to stable simulations that reproduce its climate, and the use of an RF allowed physical constraints, such as energy conservation and non-negative surface precipitation, to be respected.
More recently, \citet{yuval2020stable} (hereinafter referred to as YOG20)  learned an RF parameterization from output of a three-dimensional
high-resolution idealized atmospheric model. The parameterization led to stable simulations that replicate the climate of the high-resolution simulations at several coarse resolutions. 


Despite the success of RFs in these cases, NNs have some computational advantages over RFs such as the possibility of greater accuracy and needing substantially less memory when implemented.  
Furthermore, an NN parameterization could potentially be implemented at reduced precision in 
GPUs, TPUs and even in CPUs \citep{vanhoucke2011improving}
leading to very efficient parameterizations. 
For example, modern hardware developments that use half precision (16-bit)
arithmetic on recent generation NVidia A100 GPU devices \citep{Nvidia_2020} allow up to 64
times the peak compute performance (0.6 PFlop/s at 16-bit precision), 
compared to 64-bit arithmetic (9.7 TFlop/s at
64-bit precision). Although these are theoretical peak numbers, a
first exascale application in Earth science has already been realized
for image classification based on GPU technology  with 16-bit arithmethic \citep{kurth2018exascale}.
Google TPUv3 devices \citep{jouppi2020domain}
 have similarly optimized capability (0.12 PFlop/s at 16-bit
precision), for reduced precision floating point arithmetic to support
deep learning applications.
Furthermore, previous studies have found that using deep NNs with half-precision 
multipliers has little effect on accuracy when used for classification 
 \citep{courbariaux2014training,miyashita2016convolutional,micikevicius2017mixed,das2018mixed}. 
For regression tasks, which are needed for parameterizations, a reduced-precision NN might not be as accurate as a full-precision NN, but it has been argued
that the dynamics at small horizontal length scales represented by parameterizations does not need the same level of precision as for large-scale dynamics because of the unpredictable and stochastic nature of the small-scale flow  \citep{palmer2014more}. Indeed,  use of reduced precision in parts of weather or climate models has already been proposed as a way to increase the speed of simulations and reduce their energy cost \citep{duben2014use,duben2014benchmark,duben2017study}.


However, there have been issues of instability and climate drift when 
NN parameterizations are used in GCMs.
For example, \citet{rasp2018deep} used NN parameterizations
in the super-parametrized community atmosphere model (SPCAM) 
and found they are often
unstable when coupled dynamically to a GCM. 
\citet{rasp2018deep} were able 
to acheive a stable simulation with an NN parameterization, but they found that small changes to the configuration led to blow ups in the simulations \citep{rasp2019online}. Furthermore, when they quadrupled the number of embedded cloud-resolving columns (from 8 to 32) within each coarse-grid cell of SPCAM they found that instabilities returned \citep{brenowitz2020interpreting}.
\citet{brenowitz2018prognostic,brenowitz2019spatially} 
learned an NN parameterization in a near-global cloud system 
resolving model (CRM) and were able to deal with instabilities
by removing the upper-atmospheric humidity and temperature from the input space and by using a training cost function that takes into account the predictions from several forward time steps. 
Although these changes in the learning structure led to stable simulations
at coarse resolution with the NN parameterization,
 the climates of these simulations drifted on longer time scales and were not accurate. 
\citet{brenowitz2020interpreting} found using spectral
analysis that instability occurs when GCMs are coupled to NNs that support unstable gravity
waves with certain phase speeds. 
A study by \citet{ott2020fortran} tested the stability of simulations coupled to more than a hundred different NNs, and found a correlation between offline performance (i.e., the quality of predictions from NNs when they are not coupled to a GCM) and how long simulations run before they blow up. These results suggest that an exhaustive hyperparameter tuning might be necessary in order to achieve stability in GCM simulations that are coupled to NNs.

RFs might be more stable than NNs since their predictions for any given input are averages over samples in the training dataset, and thus they can automatically satisfy physical constraints such as energy conservation  \cite[YOG20]{Ogorman2018using} and make conservative predictions for samples outside of the training data (NNs can also be forced to satisfy analytic constraints \citep{beucler2019achieving}, but such NNs have not yet been coupled to a GCM).
However, the RFs and NNs in the studies mentioned above used different training datasets and preprocessed the high-resolution data differently. 
Therefore, it is difficult to determine if the stability arises from 
the different processing of the high-resolution model output
or due to the the different properties of RFs compared to NNs. 
The two main differences in the data processing between 
YOG20 and the NNs studies mentioned above are that (1) the predicted tendencies were calculated
accurately for the instantaneous atmospheric state (YOG20) rather than approximating them based on differences over 3-hour periods \citep{brenowitz2018prognostic,brenowitz2019spatially} and that (2) the subgrid corrections were calculated independently for each physical process (YOG20) rather than for the combined effect of several processes  \citep{rasp2018deep,brenowitz2018prognostic,brenowitz2019spatially}. 


Here we learn an NN parameterization using  
the same data processing that was used to learn an RF parameterization in YOG20.
We show that 
implementing this parameterization in the same model 
at coarse resolution leads to stable 
simulations with a climatology 
that resembles the one obtained from a high-resolution simulation, 
and we compare the performance of 
an NN parameterization to the performance 
of an RF parameterization.
We also test how the simulated climate is affected 
when reducing the precision of the inputs  and outputs of the NN parameterization.

We first describe the high-resolution simulation from which the training data was calculated and how this  data was coarse-grained (section~\ref{sec:methonds}).
We then describe the structure of the NN parameterization and 
explain how this structure ensures that NN parameterization is consistent with several physical properties (section~\ref{sec:NN parameterization explained}). 
We compare simulations using the NN parameterization to the high-resolution simulation  and to a simulation with an RF parameterization, and we investigate the dependence of climate  on the numerical precision of the NN parameterization (section~\ref{sec:results}). 
Finally, we give our conclusions (section~\ref{sec:conclusions}).

\section{Methods \label{sec:methonds}}

\subsection{Simulations \label{sub:hi-res simulation}}
All simulations were run on
a quasi-global aquaplanet configured as an equatorial beta-plane
using the System for Atmospheric Modeling
(SAM) version 6.3 \citep{khairoutdinov2003cloud}.
The domain has
zonal width of $6,912$km and meridional extent of $17,280$km.
The distribution of sea surface temperature (SST) is specified to be
the \textcolor{black}{qobs} SST distribution \citep{neale2000standard}, which is 
zonally and hemispherically symmetric.
There are 48 vertical levels.
We use hypohydrostatic rescaling of the equations of motion with a scaling
factor of 4, which increases the horizontal length scale of convection and allows us to use a horizontal
grid spacing of 12km for the high-resolution simulation (referred to as hi-res)
while explicitly representing both convection and planetary scale circulations in the same quasi-global simulation
\citep{kuang2005new,pauluis2006sensitivity,garner2007resolving,boos2016convective,fedorov2019tropical}. 
The hi-res simulation is the same simulation that was used for training in YOG20.
It was recently shown that the tropical distributions of precipitation intensity and precipitation cluster size in hi-res are similar to the distributions calculated from satellite-based observations  \citep{ogorman2020_prcip}. 
Further details of the model configuration are given in YOG20.

In addition to hi-res, 
we also consider simulations at horizontal grid spacings of 96km and 192km, which will be referred to
as x8 and x16, respectively, since they correspond to coarser grid spacings by 
factors of 8 and 16, respectively. 
We ran a simulation at 96km horizontal grid spacing without an NN parameterization (x8), several simulations at 96km horizontal grid spacing with an NN parameterization (x8-NN, variants of this simulation are described in the text below), and simulations at 192km horizontal grid spacing with (x16-NN) and without (x16) an NN parameterization.
All simulations were run for $600$ days. The first $100$ days are considered as spin up,
and the presented results are taken from the last $500$ days of each simulation.
Simulations with the NN parameterization start with initial conditions taken from the last time step of the simulations without the NN parameterization at the same resolution.

\subsection{Coarse-graining and calculation of subgrid terms \label{sub:coarse-graining}}
The NN parameterization aims to account for unresolved
processes that  act in the vertical and affect thermodynamic and moisture prognostic variables.
There are three thermodynamic and moisture prognostic variables in SAM \citep{khairoutdinov2003cloud}:
liquid-ice static energy $h_{\rm{L}}$,
total non-precipitating water mixing ratio $q_{\rm{T}}$,
and  precipitating water mixing ratio $q_{\rm{p}}$.
Since  $q_{\rm{p}}$ is a variable that varies on short time scales not typically resolved in climate models, we do not include $q_{\rm{p}}$ in the NN parameterization scheme
 following the ``alternative'' parameterization approach in YOG20. 
 Consequently,  we reformulate the  equations of motion, and define a different thermodynamic variable ($H_{\rm{L}}$) that does not include the energy associated with precipitating water (text S1).

For each 3-hourly snapshot from hi-res, we coarse grain the prognostic variables ($u,v,w,H_{\rm{L}},q_{\rm{T}},q_{\rm{p}}$, where $u,v,w$ are the zonal, meridional and vertical wind, respectively), vertical advective fluxes, sedimentation fluxes, surface fluxes, tendency of non-precipitating water mixing ratio due to cloud microphysics, turbulent diffusivity, radiative heating and temperature. Coarse-graining is performed by spatial averaging as in YOG20  to horizontal grid spacings of 96km (x8) and 192km (x16). 
 
We define the resolved flux of a given variable as the flux calculated using the dynamics and physics of SAM with the coarse-grained prognostic variables as inputs.
The flux due to unresolved (subgrid) physical processes is calculated as the difference between the coarse-grained flux and the resolved flux. 
For example, the subgrid flux of $H_{\rm{L}}$  due to vertical advection is calculated as
\begin{linenomath*}
\begin{equation}
\left({H}_{\rm{L}}\right)_{\rm{adv}}^{\rm{subg-flux}} =\left( \overline{w H_{\rm{L}}}- \overline{w} \overline{H}_{\rm{L}}\right),
\label{eq:sgs adv calc}
\end{equation}
\end{linenomath*}
where overbars denoted coarse-grained variables. For each high-resolution snapshot the coarse-grained prognostic fields are used to 
calculate the resolved vertical advective, sedimentation and surface fluxes. The subgrid fluxes are then calculated by taking the difference between the coarse-grained and resolved fluxes.




\section{Neural network parameterization \label{sec:NN parameterization explained}}
\subsection{Parameterization structure \label{sub:nn_structure}}
The structure of the NN parameterization is broadly similar to the RF parameterization used in YOG20 except 
that where possible we predict fluxes and sources and sinks (rather than net tendencies in YOG20) 
in order to incorporate physical constraints into the NN parameterization (section \ref{sub:Consistent modelling}). 
By contrast, for the RF parameterization in YOG20 it was sufficient to just predict net tendencies 
because the RF predicted averages over 
subsamples of the training dataset and thus automatically 
respected physical constraints such as energy conservation.

\begin{figure}
\centerline{\includegraphics[scale=0.75]{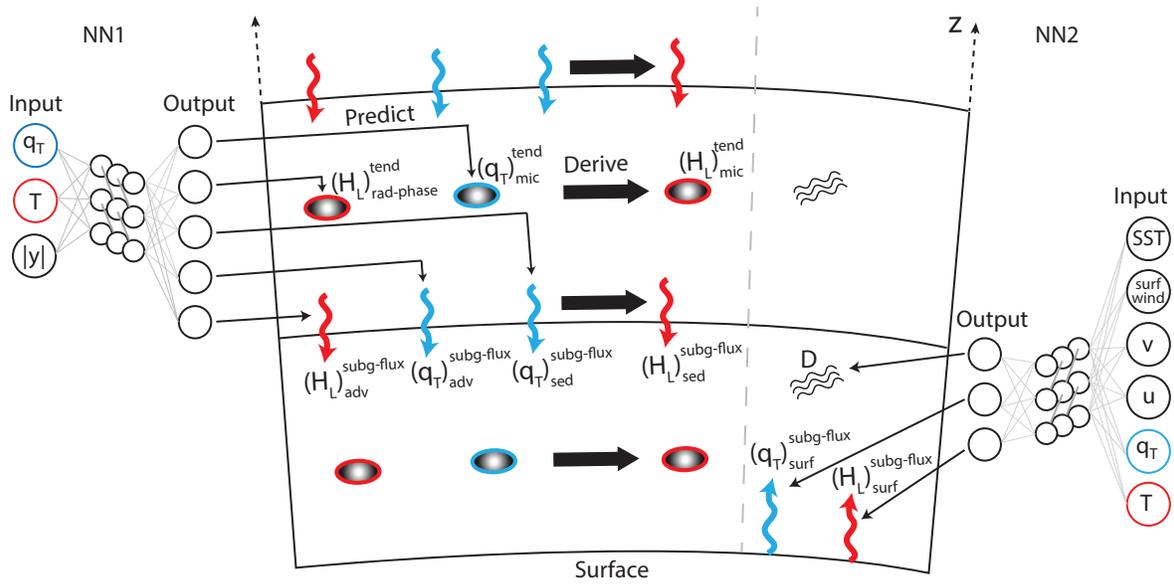}}
\caption{Illustration of the parameterization structure. 
There are two different NNs included in the parameterization, each having its own inputs and outputs (full description in section~\ref{sub:nn_structure}). Arrows indicate subgrid fluxes due to vertical advection and sedimentation, and ellipses indicate tendencies associated with sources/sinks due to cloud microphysics, radiation and phase changes of precipitating water. Blue (red) color indicates a variable, tendency or flux associated with moisture (energy). 
\textbf{}}
\label{fig:NN_caricature}
\end{figure}

The NN parameterization is composed of two different NNs (Figure~\ref{fig:NN_caricature}). 
The first NN, referred to as NN1, predicts the vertical profiles of the subgrid vertical advective fluxes of $H_{\rm{L}}$ ($(H_{\rm{L}})_{\rm{adv}}^{\rm{subg-flux}}$) and $q_{\rm{T}}$ ($(q_{\rm{T}})_{\rm{adv}}^{\rm{subg-flux}}$), the subgrid flux due to cloud ice sedimentation ($(q_{\rm{T}})_{\rm{sed}}^{\rm{subg-flux}}$), the coarse-grained  tendency of $q_{\rm{T}}$ due to cloud microphysics ($(q_{\rm{T}})_{\rm{mic}}^{\rm{tend}}$), and the sum of the total radiative heating and the heating from phase changes of precipitation ($(H_{\rm{L}})_{\rm{rad-phase}}^{\rm{tend}}$, text S1). 
 Thus the outputs of NN1 are 
\begin{equation}
Y_{\rm{NN1}} = ((H_{\rm{L}})_{\rm{adv}}^{\rm{subg-flux}}, (q_{\rm{T}})_{\rm{adv}}^{\rm{subg-flux}}, 
(q_{\rm{T}})_{\rm{sed}}^{\rm{subg-flux}}, (q_{\rm{T}})_{\rm{mic}}^{\rm{tend}}, (H_{\rm{L}})_{\rm{rad-phase}}^{\rm{tend}}),
\label{eq:NN-tend_outputs}
\end{equation}
where the superscript subg-flux refers to subgrid fluxes and the superscript tend refers to the total tendency due to a process. 
The tendencies due to cloud microphysics, radiative heating and heating from phase changes of precipitation are treated as entirely subgrid. In the case of cloud microphysics and phase changes of precipitation, this is because it is not possible to calculate the resolved values of these processes when $q_{\rm{p}}$ is not used as a prognostic variable in the simulations (text S1).
Tendencies are predicted at the lowest 30 ``full'' model levels (below $z=13.9$km), while subgrid ice sedimentation fluxes are predicted at the lowest 30 ``half'' model levels, and vertical advective fluxes are predicted at the 29 ``half'' model levels above the surface (since advective fluxes is zero at the surface over ocean). Thus, NN1 has $29 \times 2 + 30 \times 3=148$ outputs. 
We do not use NN1 to predict outputs for levels above $13.9$km ($\approx 134$hPa) since 
the NN parameterization is not accurate at these levels, we want to avoid predicting near the sponge layer which is active at heights above $20$km, and the predicted tendencies and subgrid fluxes are small above $13.9$km with the exception of radiative heating.
Above 13.9km the subgrid fluxes and microphysical tendency are set to zero and 
the radiative heating tendencies calculated at coarse resolution from SAM are used.

The inputs for NN1 are the resolved vertical profiles of $q_{\rm{T}}$ and temperature ($T$), as well as the distance to the equator ($|y|$, which is a proxy for insolation and surface albedo), giving $30 \times 2 +1=61$ inputs. We verified that using top of atmosphere insolation as input instead of the distance to equator does not change the results presented in this study (Figure~S1). 
We found that using all $48$ levels of $T$ and $q_{\rm{T}}$ as inputs in NN1 leads to an instability, possibly related to instabilities found 
previous studies when $q_{\rm{p}}$ is not used a prognostic variable \cite[YOG20]{brenowitz2019spatially,brenowitz2020interpreting}. However, 
here we include inputs up to the lower stratosphere while \citet{brenowitz2019spatially} and \citet{brenowitz2020interpreting} removed some inputs above the mid-troposphere to achieve stability, whereas here it is sufficient to not use inputs in the stratosphere.

The second NN, referred to as NN2, predicts
subgrid surface fluxes of $H_{\rm{L}}$ and $q_{\rm{T}}$ and the 
coarse-grained vertical turbulent
diffusivity ($\overline{D}$) that is used for $H_{\rm{L}}$  and $q_{\rm{T}}$.
We only predict $\overline{D}$ in the lower troposphere
(the 15 model levels below $5.7$km)
because it decreases in magnitude with height (YOG20), and the diffusivity calculated at coarse resolution from SAM is used above $5.7$km.
Hence the outputs 
 of NN2 are 
 \begin{equation}
 Y_{\rm{NN2}} = (\overline{D}, (h_{\rm{L}})_{\rm{surf}}^{\rm{subg-flux}},(q_{\rm{T}})_{\rm{surf}}^{\rm{subg-flux}}),
 \end{equation}
giving $15+1+1=17$ outputs.
The inputs of NN2 are chosen to be the lower tropospheric vertical profiles of $T$, $q_{\rm{T}}$, $u$,  $v$,
surface wind speed (wind$_{\rm{surf}}$) and SST, so that
$X_{\rm{NN2}} = (T,q_{\rm{T}},u,v,\rm{wind_{\rm{surf}}},SST)$, giving
$4\times15+1 + 1 =62$ features,
where $v$ in the southern hemisphere is multiplied 
by $-1$ when used as an input 
(see further discussion in YOG20). 

The tendencies due to subgrid vertical advective, sedimentation and surface fluxes are calculated online (i.e., when running SAM  with the parameterization) from the predicted fluxes. The subgrid energy flux due to ice sedimentation is also calculated online as
\begin{equation}
(H_{\rm{L}})_{\rm{sed}}^{\rm{subg-flux}} = -L_{\rm{s}} (q_{\rm{T}})_{\rm{sed}}^{\rm{subg-flux}},
\label{eq:H_sed_flux}
\end{equation}
where $L_{\rm{s}}$ is the latent heat of sublimation. 
Similarly, the tendency of energy due to cloud microphysics is calculated online as 
\begin{equation}
(H_{\rm{L}})_{\rm{mic}}^{\rm{tend}} =-L_{\rm{p}} (q_{\rm{T}})_{\rm{mic}}^{\rm{tend}},
\label{eq:H_mic_tend}
\end{equation}
where $L_{\rm{p}} = L_{\rm{c}} + L_{\rm{f}}(1-\omega_{\rm{p}})$ is the effective latent heat associated with precipitation ($L_{\rm{c}}$ and $L_{\rm{f}}$ are the latent heat of condensation and fusion, respectively, and $\omega_{\rm{p}}$ is the precipitation partition function determining the ratio between ice and liquid phases).

The results presented in the main paper are for NNs (both NN1 and NN2) with five 
layers (four hidden layers) with 128  nodes 
and rectified linear unit (ReLu) activation functions. 
Results for different NN architectures are shown in the supplementary information (Figure~S2).
\textcolor{black}{More details about the train and test datasets, the NNs training protocol 
and how inputs and outputs are rescaled can be found in the supplementary information (text S2)}

When running simulations  with the NN parameterization,  we limit the predictions of $q_{\rm{T}}$ fluxes and tendencies such that we prevent $q_{\rm{T}}$ from being assigned with negative values.  More specifically, if the NN parameterization predicts any flux or any tendency at any vertical level that would lead to negative $q_{\rm{T}}$ values if it acted on its own, we reduce this tendency or flux magnitude such that the minimal value that can be assigned to $q_{\rm{T}}$ is zero. 
Removing this constraint (and allowing  $q_{\rm{T}}$ to be assigned with negative values) leads to a substantially different climate when the NN parameterization is used, although the simulations remain stable.

\subsection{Physical consistency of the parameterization \label{sub:Consistent modelling}}

Previous studies that used NN parameterizations usually predicted 
the sum of tendencies due to several different processes as a single output
\citep{gentine2018could,rasp2018deep,brenowitz2018prognostic,brenowitz2019spatially}.
The coarse-graining and calculation of subgrid terms
 that is used in this study (section \ref{sub:coarse-graining}) 
enables us to predict the effect of each process on the prognostic variables separately (section \ref{sub:nn_structure}),
 and where possible, the effect is diagnosed from other predicted outputs 
 (equations~\ref{eq:H_sed_flux} and \ref{eq:H_mic_tend}). 
 Furthermore, the NN parameterization predicts 
fluxes instead of tendencies where possible. 
These differences make the NN parameterization presented 
in this study physically consistent in the following ways:
\begin{enumerate}[label=\alph*]
\item Predicting the subgrid fluxes due to vertical advection instead of the subgrid tendencies guarantees energy and water are conserved by these fluxes. Similarly, predicting the flux due to sedimentation guarantees that changes in the energy of the atmospheric column due to sedimentation are only due to sedimentation that reaches the surface.
\item Changes in $H_{\rm{L}}$ due to cloud microphysics and ice sedimentation are not predicted by the NN, but are instead diagnosed from equations~\ref{eq:H_sed_flux} and \ref{eq:H_mic_tend} (Figure~\ref{fig:NN_caricature}).
Diagnosing these tendencies guarantees that the sources or sinks of energy at each grid point due to cloud microphysics and sedimentation are consistent with the amount of moisture that was subtracted or added at that grid point. 
\item The NN predicts the coarse-grained vertical diffusivity (instead of predicting subgrid diffusive tendencies or fluxes) which ensures that diffusive fluxes are downgradient and conserve energy and water in each atmospheric column.
\item The precipitation 
is diagnosed from the NN outputs (text~S1).
Diagnosing the precipitation was not done in some previous studies that used NN parameterization in which the NN was used to predict precipitation directly \citep{rasp2018deep} or the NN outputs could only be used to predict the difference between precipitation minus evaporation \citep{brenowitz2019spatially}.
\end{enumerate}

Properties a,b, and c ensure that the NN parameterization exactly conserves energy in the sense that the column integrated frozen moist static energy is conserved in the absence of radiative heating, heating from phase changes of precipitation, surface fluxes, and conversion of condensate to graupel or snow (see Text S3).

\section{Results \label{sec:results}}
\subsection{Simulation with neural-network parameterization \label{sub:online sim}}
\begin{figure}
\centerline{\includegraphics[scale=0.8]{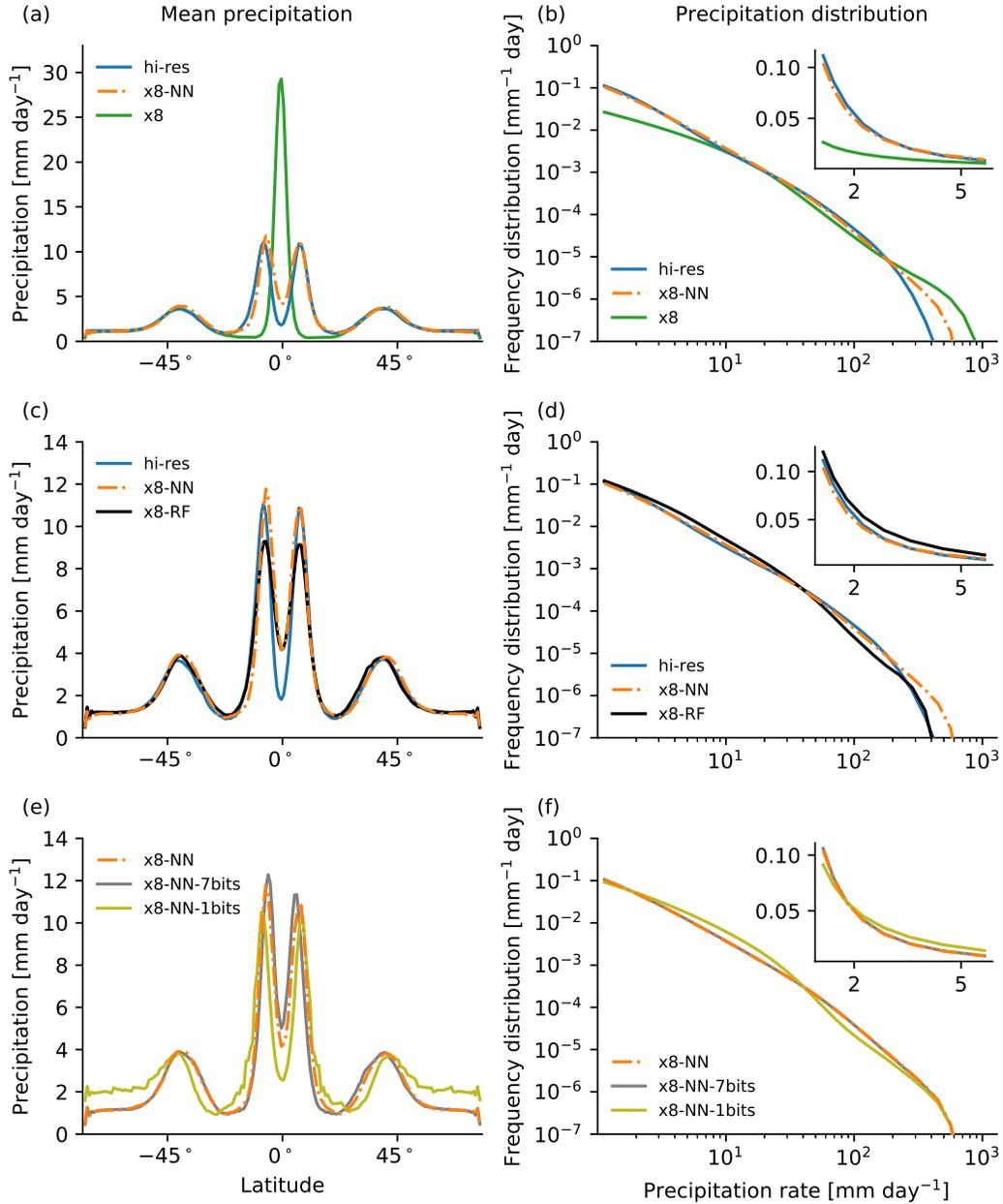}}
\caption{ 
(a,c,e) Zonal- and time-mean precipitation rate as a function of latitude and (b,d,f) frequency distribution of 3-hourly precipitation rate.
Results are shown for 
the high-resolution simulation (hi-res in blue; panels a-d), the coarse-resolution simulation (x8 in green; panels a-b), the coarse-resolution simulation with the NN parameterization (x8-NN in orange dash-dotted; panels a-f), the coarse-resolution simulation with the RF parameterization (x8-RF in  black; panels c-d), 
and simulations with reduced numerical precision of the inputs and outputs of the NN parameterization with
7 significant bits in the mantissa (x8-NN-7bits in gray; panels e-f),
and 1 significant bit in the mantissa (x8-NN-1bits in yellow; panels e-f)  as compared to 23 bits in the mantissa for x8-NN.
The frequency distribution is shown for axes with linear scale in the insets of b,d,f. }
\label{fig:precip_online_x8_all}
\end{figure}

%
To assess the NN parameterization,
 we compare the climates of  x8, x8-NN
 (using NNs with five fully connected layers, text~S2) 
 and the hi-res. 
We focus on the zonal- and time-mean 
precipitation distribution (Figure~\ref{fig:precip_online_x8_all}a) and the 
frequency distribution of the 3-hourly precipitation rate (Figure~\ref{fig:precip_online_x8_all}b).
The frequency distribution is known to be sensitive to subgrid 
parameterization of moist convection \citep{wilcox2007frequency}, 
and the latitudinal distribution of mean precipitation is
especially sensitive to subgrid parameterizations in the
zonally and hemispherically symmetric 
aquaplanet configuration used here \citep{mobis2012factors}. 
The frequency distribution is estimated using $34$ 
bins that are equally spaced in the logarithm of precipitation rate, where
the lowest bin starts at $1 \rm{mm~ day^{-1}}$, and the distribution 
is normalized such that it integrates to one when considering the whole distribution 
(including values lower than $1 \rm{mm~ day^{-1}}$).
The hi-res and the x8-NN simulations exhibits a similar double ITCZ structure, 
both in amplitude and in the latitudinal structure (Figure~\ref{fig:precip_online_x8_all}a). Note that the presence of a double ITCZ in hi-res is likely to be dependent on the exact domain and SST distribution.
In contrast, the x8 simulation exhibits a single ITCZ (Figure~\ref{fig:precip_online_x8_all}a), 
highlighting the sensitivity of the tropical circulation to changes in the 
horizontal resolution and to the inclusion of the NN parameterization. 
Also the frequency distributions of precipitation in the x8-NN
closely matches that of hi-res ($R^2 = 0.99$  as measured across bins), although x8-NN overestimates the extreme events,
while the frequency distribution in x8 differ substantially from the distribution of hi-res ($R^2 = 0.35$)
especially for weak and extreme precipitation events (Figure~\ref{fig:precip_online_x8_all}b). 
The NN parameterization also brings the zonal- and time-means of 
the zonal wind, meridional wind, eddy kinetic and 
non-precipitating water closer to the hi-res climatology, 
and outperforms the x8 simulation for these variables \textcolor{black}{(table~S1)}.
We find that these results are reproduced 
when the NN1 architecture is changed to 
have only three or four layers (Figure~S2, although the amplitude of the 
equatorial minimum of precipitation slightly varies between simulations). 
All the simulations we ran were stable and without climate drift, 
and even a substantially less accurate NN parameterization 
with two layers for NN1 (table~S2) leads to stable simulations,
though it does not capture the precipitation distributions (Figure~S2).
When training an NN 
parameterizations for a coarse-graining
factor of x16 and coupling it to a simulation with
the corresponding grid spacing, similar results are obtained and 
precipitation distributions are captured correctly \textcolor{black}{(Figure~S3)}. 
Overall, we find that coupling an NN parameterization to a 
simulation at coarse-resolution leads to 
a climate that is similar to the hi-res climate, 
and that the stability of simulations is not sensitive to the NN 
architecture, hyperparameter tuning or 
the horizontal resolution of the simulations. 

Several approximations where made when deriving the instantaneous precipitation rate 
(text S1). These approximations and inaccuracies in the NN predictions
result in negative 3-hourly surface precipitation in $20\%$ of samples in the x8-NN simulation. Nevertheless, almost all of the negative values are very small, and only $0.03\%$ of samples are less than $-1 \rm{mm~ day^{-1}}$.

%

\subsection{Comparing neural network and random forest parameterizations \label{sub:nn vs rf}}

In this section we compare the offline and online performance of NN parameterization to the (``alternative'') RF parameterization that was developed in YOG20 and also did not include $q_{\rm{p}}$. 
For offline performance  (performance when not coupled to a GCM), we find that the NN parameterization outperforms the RF parameterization across all variables (Figure~S4, text~S4).
Yet, the online performance of the x8-NN and x8-RF simulations is comparable (Figure~\ref{fig:precip_online_x8_all}c,d and table~S1). 
Both x8-NN and x8-RF have a double ITCZ as in hi-res.
The x8-NN simulation better captures the frequency distribution at most precipitation rates, though for extreme events x8-RF is more accurate (Figure~\ref{fig:precip_online_x8_all}d). 
Other climatological measures such as 
mean $q_{\rm{T}}$, meridional and zonal wind 
and eddy kinetic energy are 
better captured by x8-RF  \textcolor{black}{(table~S1)}.

One advantage of NN is that it requires less memory compared to RF. For example, for x8, NN1 is 0.3 MB and NN2 is 0.2 MB when stored in netcdf format at single precision, which is $\approx 1900$ times less memory compared to the memory needed to store the RF parameterization.
Another advantage of the NN parameterization is that the x8-NN simulation requires less CPU time than the x8-RF simulation by a factor of \textcolor{black}{$1.25$}, although both require much less CPU time than hi-res (by a factor of $54$ in the case of x8-NN).

\subsection{Reduced-precision parameterization\label{sub:missing bits}}

As discussed in the introduction, NN parameterizations run at reduced 
precision could bring considerable savings in speed and energy. 
To test the effect of 
reduced precision NNs, we run simulations 
in which the scaled inputs and outputs of the NNs are reduced in precision. 
Such simulations aim to check how precise the outputs and inputs have to be to give a similar climate to the full precision.
In our default configuration, SAM and the NNs are evaluated in single precision which corresponds to 
23 bits in the mantissa. We reduce the precision by rounding to 
7,5, 3 and 1 bits in the mantissa and the resulting simulations are referred to as x8-NN-7bits, 
x8-NN-5bits, x8-NN-3bits and x8-NN-1bits, while keeping the same number of bits in the 
exponent (8 bits). Note that the 7 bits  in the mantissa corresponds to
 the bfloat16 ``brain'' floating-point (7 bits in the mantissa, 8 in the exponent and 1 to determine the sign) 
which is used in TPUs. 

 
For x8-NN-7bits and x8-NN-5bits simulations we find that 
the resulting climate is similar,
while reducing the precision to x8-NN-3bits leads to a small degradation and keeping only 1 bit 
in the mantissa clearly degrades the results (table~S1, Figure~\ref{fig:precip_online_x8_all}e,f).
Overall, we find that it is possible to reduce the precision of inputs and outputs 
even beyond bfloat16 format without substantially affecting the climate of the simulations, suggesting that NN parameterizations at reduced precision could bring substantial advantages in speed and power requirements.

\section{Conclusions \label{sec:conclusions}}

In this study we show that an NN parameterization 
with a physically motivated structure learned 
from coarse-grained output of a three-dimensional high-resolution simulation 
of the atmosphere 
can be dynamically coupled to 
the atmospheric model at coarse resolution to give stable simulations
with a climate that is close to that of the high-resolution 
simulation. 
In contrast to the approach in \citet{rasp2018deep}  
we find that simulations with NN parameterization 
are stable for a variety of configurations, 
and in contrast to \citet{brenowitz2018prognostic,brenowitz2019spatially} they do not exhibit climate drift.
Furthermore, in contrast to \citet{ott2020fortran} we find that achieving stable simulations
does not require the NN parameterization to be very accurate in an offline test,  
and a mediocre performing NN1 with two layers (i.e., one hidden layer)
is stable when coupled to SAM. 



We use the same high-resolution model output for training that was used by
YOG20, which suggests that the stability of simulations
with an RF parameterization 
in previous studies \cite[YOG20]{Ogorman2018using}
is not only possible with RFs since
we find NNs to be robustly stable as well.
Instead, the stability of simulations with NN parameterizations
may require accurate processing of the high-resolution model output to 
obtain subgrid tendencies and fluxes.
(in addition to not including stratospheric levels). The main differences in the processing used here compared to previous 
studies with NN parameterizations of atmospheric subgrid processes are that the contribution of subgrid terms 
were calculated using the equations of the model 
for the instantaneous state of the atmosphere rather than 
approximating them based on differences over 3-hour periods \citep{brenowitz2018prognostic,brenowitz2019spatially}
and that subgrid corrections were calculated
independently for each physical process. The latter allows us
to encapsulate more physics in the parameterization,
such as by calculating fluxes and sinks rather than net tendencies. 
A direct comparison between an RF and NN parameterizations 
shows that although NNs are more accurate in offline tests, 
when coupling the parameterizations to the atmospheric model
at coarse resolution, both parameterizations have comparable results. 
Overall, our results imply that careful and accurate processing of the high-resolution output
that is used for the training may be more important than 
intensive hyperparameter tuning of an ML algorithm. 
The time step in our coarse-resolution simulations cannot be larger than roughly 75 seconds because of the explicit time stepping of turbulent diffusion in SAM
, and future work is needed to extend these results to longer time steps and to simulations with land and topography.


Finally we show that reducing the numerical 
precision of the inputs and outputs of the 
NN parameterization to bfloat16 floating-point format
leads to a similar climate compared to using single precision. 
This implies that NN parameterizations with reduced precision could be used for faster training, 
and more importantly, for reducing the computational resources  and energy 
needed to run climate simulations.
To further investigate the feasibility of NN parameterizations with reduced precision,
future work should also test NN parameterizations that were trained with reduced precision, such that the weights, biases and multiplications used during forward propagation of the NN are performed at reduced precision.
Our results also suggest that the simulated climate may not be strongly affected by reducing the precision of conventional parameterizations or super parameterizations, but in those cases it would likely be necessary to check for each particular parameterization which parts of the algorithm can be safely reduced in precision \citep{duben2017study}.

\subsubsection*{Acknowledgments}
We thank Bill Boos for providing the output from the high-resolution simulation, and Tamar Regev for help in creating figure 1 in this manuscript.
We acknowledge high-performance computing support from Cheyenne (doi:10.5065/D6RX99HX) provided by NCAR's Computational and Information Systems Laboratory, sponsored by the National Science Foundation.
We acknowledge support from the EAPS Houghton-Lorenz postdoctoral fellowship and NSF AGS-1552195.

\subsubsection*{Code availability}
Associated code, processed data from simulations with neural network and random forest parameterizations, trained neural network parameterizations and (a link to) the output of the high-resolution simulation are available at zenodo.org (doi:10.5281/zenodo.4118346).


%
%

\bibliography{yanibib_long_names}

%
%
%
%
%

\end{document}


\date{}
\maketitle

\renewcommand{\thepage}{S\arabic{page}} 
\renewcommand{\thesection}{S\arabic{section}}  
\renewcommand{\thetable}{S\arabic{table}}  
\renewcommand{\thefigure}{S\arabic{figure}}
\renewcommand{\theequation}{S\arabic{equation}}

%
%

%
%





%
%

\noindent\textbf{Contents of this file}
\begin{enumerate}
\item Text S1 to S4
\item Figures S1 to S4
\item Tables S1 to S2

Here we present further information on the equations of motion for thermodynamic and moisture variables
used in SAM and how these equations are modified when we omit precipitating water as a prognostic variable  following the approach in \citet{yuval2020stable} (hereinafter referred to as YOG20, text S1). We also describe the training data, training procedure and architecture of the neural networks (NNs) and their implementation in SAM (text S2), we show that the NN parameterization conserves the frozen moist static energy (text S3), and we compare the offline performance of the NN and RF parameterizations (text S4).

\end{enumerate}




%


\noindent\textbf{Text S1. Equations of motion used for thermodynamic and moisture variables}
The equations for these variables in SAM may be written as \citep{khairoutdinov2003cloud}
\renewcommand{\theequation}{S\arabic{equation}}
 \begin{equation}
 \frac{\partial h_{\rm{L}}}{\partial t} = -\frac{1}{\rho_{\rm{0}}} \frac{\partial}{\partial x_i}(\rho_{\rm{0}} u_i h_{\rm{L}} + F_{h_{\rm{L}} i}) -\frac{1}{\rho_{\rm{0}}}  \frac{\partial}{\partial z}(L_{\rm{p}} P_{\rm{tot}}+ L_{\rm{s}} S) +(h_{\rm{L}})_{\rm{rad}}^{\rm{tend}},
 \label{eq:hL equation} 
 \end{equation} 
\renewcommand{\theequation}{S\arabic{equation}}
\begin{equation}
\frac{\partial q_{\rm{T}}}{\partial t} = -\frac{1}{\rho_{\rm{0}}} \frac{\partial}{\partial x_i}(\rho_{\rm{0}} u_i q_{\rm{T}} + F_{q_{\rm{T}} i}) +\frac{1}{\rho_{\rm{0}}}  \frac{\partial}{\partial z}(S)  + (q_{\rm{T}})_{\rm{mic}}^{\rm{tend}},
 \label{eq:qT equation} 
\end{equation}
\renewcommand{\theequation}{S\arabic{equation}}
\begin{equation}
\frac{\partial q_{\rm{p}}}{\partial t} = -\frac{1}{\rho_{\rm{0}}} \frac{\partial}{\partial x_i}(\rho_{\rm{0}} u_i q_{\rm{p}} + F_{q_{\rm{p}} i}) + \frac{1}{\rho_{\rm{0}}}  \frac{\partial}{\partial z}( P_{\rm{tot}})-(q_{\rm{T}})_{\rm{mic}}^{\rm{tend}},
 \label{eq:qp equation}
\end{equation}
where  
$h_{\rm{L}}$ is the liquid/ice water static energy ($= c_{\rm{p}}T + gz - L_{\rm{c}}(q_c + q_{\rm{r}}) - L_{\rm{s}}(q_{\rm{i}} + q_{\rm{s}} + q_{\rm{g}})$, and 
$q_{\rm{c}}$, $q_{\rm{i}}$, $q_{\rm{r}}$, $q_{\rm{s}}$ and  $q_{\rm{g}}$ are mixing ratios of 
cloud water, cloud ice, rain, snow and graupel, respectively);
$q_{\rm{p}}$ is the total precipitating water (rain + graupel + snow) mixing ratio;  
$q_{\rm{T}}$ is the non-precipitating water (water vapor + cloud water + cloud ice) mixing ratio;
$\rho_{\rm{0}}(z)$ \textcolor{black}{is the reference density profile}; 
$u_i = (u,v,w)$ is the three-dimensional wind; 
$F_{Bi}$ is the diffusive flux of variable $B$;
$P_{\rm{tot}}$ is the precipitation mass flux (due to rain, graupel and snow); 
$S$ is the sedimentation mass flux; 
$(h_{\rm{L}})_{\rm{rad}}^{\rm{tend}}$ is the tendency due to radiative heating; 
$ (q_{\rm{T}})_{\rm{mic}}^{\rm{tend}}$ is the non-precipitating water mixing ratio tendency due
to autoconversion,  collection, aggregation, evaporation and sublimation of precipitation.
$L_{\rm{c}}$, $L_{\rm{f}}$ and $L_{\rm{s}}$ are the latent heat of condensation, fusion and sublimation, respectively; 
$L_{\rm{p}} = L_{\rm{c}} + L_{\rm{f}}(1-\omega_{\rm{p}})$ is the effective latent heat associated with precipitation, where $\omega_{\rm{p}}$ is the precipitation partition function determining the ratio between ice and liquid phases.

The precipitating water mixing ratio $q_{\rm{p}}$ is a variable that varies on short time scales and is often not used as a prognostic variable in climate models, which typically have a coarser grid and larger time step than the high-resolution simulation. Therefore, we do not include $q_{\rm{p}}$ in the NN parameterization scheme that is presented in this study to make our results more generally applicable. 
Following YOG20, we define a new prognostic energy variable ($H_{\rm{L}}$) that does not include the precipitating water ($q_{\rm{p}}$):
\renewcommand{\theequation}{S\arabic{equation}}
\begin{equation}
H_{\rm{L}} = c_{\rm{p}}T + gz - L_{\rm{c}} q_{\rm{c}} - L_{\rm{s}} q_{\rm{i}},
\end{equation}
Neglecting variations of  $L_{\rm{p}}$  in the horizontal and in time (which are small compared to the vertical variations in $L_{\rm{p}}$), allows us to write the prognostic equation for $H_{\rm{L}}$ as 
\renewcommand{\theequation}{S\arabic{equation}}
\begin{align}
\frac{\partial H_{\rm{L}}}{\partial t} &= -\frac{1}{\rho_{\rm{0}}} \frac{\partial}{\partial x_i}(\rho_{\rm{0}} u_i H_{\rm{L}} ) 
-\frac{1}{\rho_{\rm{0}}}  \frac{\partial}{\partial z}(L_{\rm{s}} S)
-L_{\rm{p}}(q_{\rm{T}})_{\rm{mic}}^{\rm{tend}} + (h_{\rm{L}})_{\rm{rad}}^{\rm{tend}}   \nonumber \\
&  - \frac{1}{\rho_{\rm{0}}} \frac{\partial F_{H_{\rm{L}} i}}{\partial x_i}
 + \frac{1}{\rho_{\rm{0}}} \frac{\partial L_{\rm{p}}}{\partial z} (\rho_{\rm{0}} w q_{\rm{p}} + F_{q_{\rm{p}} z}  -  P_{\rm{tot}}) \label{eq_HL_tend}
\end{align}
where $F_{H_{\rm{L}} i} =F_{h_{\rm{L}} i} +  L_{\rm{p}} F_{q_{\rm{p}} i}$, and the last term on the right hand side results from heating due to phase changes of precipitation. We define the sum of the total radiative heating and the heating from phase changes of precipitation as $(H_{\rm{L}})_{\rm{rad-phase}}^{\rm{tend}}=(h_{\rm{L}})_{\rm{rad}}^{\rm{tend}}  + \frac{1}{\rho_{\rm{0}}} \frac{\partial L_{\rm{p}}}{\partial z} (\rho_{\rm{0}} w q_{\rm{p}} + F_{q_{\rm{p}} z}  -  P_{\rm{tot}})$.

To find an expression for surface precipitation we assume that at coarse resolution (when the NN parameterization is used) we can neglect  surface diffusive and horizontal  fluxes of $q_{\rm{p}}$ and the time derivative of $q_{\rm{p}}$ in Equation~(\textcolor{black}{\ref{eq:qp equation}}).
Using these assumptions and vertically integrating Equation~(\ref{eq:qp equation}) 
over the column gives an expression for the surface precipitation rate due to rain, graupel and snow:
\renewcommand{\theequation}{S\arabic{equation}}
\begin{equation}
P_{\rm{tot}}(z=0) = \int_{0}^{\infty} \rho_{\rm{0}} (q_{\rm{T}})_{\rm{mic}}^{\rm{tend}}  dz. 
\label{eq:precip_noqp}
\end{equation}
Following conventions used in SAM, we add any sedimentation at the surface to the surface precipitation rate and the total surface precipitation rate is calculated as:
\renewcommand{\theequation}{S\arabic{equation}}
\begin{equation}
P_{\rm{tot}}(z=0) + S(z=0) = 
\int_{0}^{\infty} \rho_{\rm{0}} (q_{\rm{T}})_{\rm{mic}}^{\rm{tend}}   dz  + (q_{\rm{T}})_{\rm{sed}}^{\rm{subg-flux}}(z=0) 
+ S^{\rm{resolved}}(z=0), 
\label{eq:precip_diagnose}
\end{equation}
where $S^{\rm{resolved}}(z=0)$ is calculated online using resolved fields in SAM.

\noindent\textbf{Text S2. Training and implementation}



The data set for the  NN parameterization is obtained from $337.5$ days of 3-hourly model output taken from the hi-res simulation. This data was split into train and test datasets, where the first $320.625$ days ($95\%$ of the data) were used for training, and the last $16.875$ days ($5\%$ of the data) were used as a test dataset.  
To easily upload all data into the RAM during the training procedure, and to decrease the correlation between different training samples for each 3-hourly snapshot that was used, 
 we reduced the training data set size for the coarse-graining factor of x$8$
by randomly subsampling
30 (out of 72) atmospheric columns at each latitude for each snapshot.   
This results in training dataset size of $13,856,040$, where 
a sample is defined as an individual
atmospheric column for a given horizontal location and time step.
When using a coarse-graining factor of x$16$, we use all $36$
longitudinal grid points at each latitude, giving  $8,313,660$ 
training samples. 
We note that the split to train and test datasets is slightly different compared to YOG20 where we used  $10\%$ of the data as a test dataset. 


The NNs training is implemented in Python using PyTorch \citep{paszke2017automatic}.
The NNs weights and biases are optimized by the Adam optimizer \citep{kingma2014adam} combined with a cyclic learning rate \citep{smith2017cyclical}. We use $1024$ samples in each batch and train over $8000$ batches before completing a full cycle in the learning rate. 
We use $12$ epochs, where the first seven epochs are trained with a minimal learning rate of $0.0002$ and a maximal learning rate of $0.002$, and five additional epochs are trained after reducing both the minimum and maximum learning rates by a factor of $10$.  The NNs are stored as netcdf files, and then implemented in SAM using a Fortran module. 
The results presented in this work are for NNs with 128 nodes at each hidden layer and rectified linear unit activations (ReLu), and unless stated differently, NNs have five densely connected layers.
Training typically takes 20 minutes when using 10 CPU cores.
We note that we do not use batch normalization \citep{ioffe2015batch} since it leads to unstable simulations. The reason that using batch normalization leads to unstable simulations is that some neurons have vanishingly small variance, which leads to divergence of prognostic fields, typically within very few time steps. 


\textcolor{black}{Prior to training, each input (feature) of both NNs and the outputs of NN2 were standardized by removing the mean and rescaling to unit variance, where for the diffusivity the mean and variance were calculated across all $15$ vertical levels.
A different approach was used for the outputs of NN1 in order to weight 
the effects of the different tendencies and fluxes consistently. 
We first define for each training sample and subgrid process 
the equivalent tendencies at all vertical levels 
associated with the fluxes by calculating tendencies
due to the predicted fluxes. We then multiplied all $q_{\rm{T}}$ 
tendencies by $L_{\rm{c}}$
such that all tendencies 
have units of $\rm{J} \ \rm{s}^{-1}$. Next, we calculated the variance of 
each of the tendencies associated with each of the
outputs 
(variance was calculated across all 30 levels of each process) 
and normalized them all by the variance of $(H_{\rm{L}})_{\rm{rad-phase}}^{\rm{tend}}$ 
which had the smallest variance.
The resulting normalized variances were 
$3.29$ for $(H_{\rm{L}})_{\rm{adv}}^{\rm{subg-flux}}$,
$22.30$ for $(q_{\rm{T}})_{\rm{adv}}^{\rm{subg-flux}}$,
$14.93$ for $ (q_{\rm{T}})_{\rm{mic}}^{\rm{tend}}$, 
$2.26$ for $(q_{\rm{T}})_{\rm{sed}}^{\rm{subg-flux}}$ and
$1.00$  for $(H_{\rm{L}})_{\rm{rad-phase}}^{\rm{tend}}$.
Finally, the outputs of NN1 were standardized by removing the mean and rescaling to the normalized variance.
In the reduced precision simulations, 
we reduce the precision of NN scaled inputs and scaled (direct) outputs, 
but we do not reduce the precision of these means and scaling factors.}

\noindent\textbf{Text S3. Conservation of frozen moist static energy}

We define the frozen moist static energy (FMSE; \citet{kuang2006mass}) as 
\renewcommand{\theequation}{S\arabic{equation}}
\begin{equation}
\mathrm{FMSE} = H_{\rm{L}} + L_{\rm{c}} q_{\rm{T}} = c_{\rm{p}}T + gz + L_{\rm{c}} q_{\rm{v}} -L_{\rm{f}} q_{\rm{i}},
\end{equation}
where $q_{\rm{v}}$ is the mixing ratio of water vapor and all other variables are defined in text~S1.
To get an evolution equation for the column integrated FMSE we multiply equation~\ref{eq:qT equation} by the latent heat of condensation, then sum the multiplied equation with equation~\ref{eq_HL_tend}, and integrate this sum in the vertical with density weighting which gives
\renewcommand{\theequation}{S\arabic{equation}}
\begin{align}
\frac{\partial}{\partial t} \int_0^{\infty} \rho_{\rm{0}} (L_{\rm{c}}q_{\rm{T}} + H_{\rm{L}}) dz &= (L_{\rm{c}} F_{q_{\rm{T}} z} + F_{H_{\rm{L}} z} )(z =0) \nonumber \\
&  +  L_{\rm{f}} S (z=0) \nonumber \\
& + \int_0^{\infty}  \rho_{\rm{0}} (H_{\rm{L}})_{\rm{rad-phase}}^{\rm{tend}} dz \nonumber \\
& -  \int_0^{\infty} \rho_{\rm{0}}  (1-\omega_{\rm{p}})L_{\rm{f}} (q_{\rm{T}})^{\rm{tend}}_{\rm{mic}} dz \nonumber \\
&-\int_0^{\infty}  L_{\rm{c}} \nabla_h \cdot  (\rho_{\rm{0}} u_h q_{\rm{T}}+F_{q_{\rm{T}} h})
+ \nabla_h \cdot (\rho_{\rm{0}} u_h H_{\rm{L}}+F_{H_{\rm{L}} h}) dz
\label{eq:FMSE_conservation}
\end{align}
where the subscript $h$ refers to the two horizontal coordinates.
The terms in the right hand side of equation~\ref{eq:FMSE_conservation} represent FMSE tendencies due to:
(1) surface fluxes,
(2) the latent heat of fusion when ice sediments reach the surface,
(3) radiative heating and heating from phase changes of precipitation,
(4) the latent heat of fusion when condensate is converted to graupel or snow, and due to
(5) horizontal advection and horizontal diffusion. 

The derivation of equation~\ref{eq:FMSE_conservation} is unaffected by the presence of the NN parameterization because: (1) the NN parameterization predicts vertical advective fluxes rather than their associated tendencies and these fluxes are set to zero at the lower boundary, (2) the NN parameterization predicts the diffusivity instead of diffusive tendencies, and (3) the NN parameterization uses equations 4 and 5 in the main paper to diagnose the subgrid fluxes or tendencies of $H_{\rm{L}}$ due to sedimentation and cloud microphysics. The NN parameterization does not affect the horizontal fluxes.
We thus conclude from equation~\ref{eq:FMSE_conservation} that the NN parameterization conserves the column integrated FMSE in the absence of radiative heating, heating from phase changes of precipitation, surface fluxes, conversion of condensate to graupel or snow and sedimentation that reaches the surface. 



\noindent\textbf{Text S4. Offline comparison between random-forest parameterization and neural-network parameterization}

Overall, the input and output structures of the NN and RF parameterizations have similarities,
though several important differences exist. 
The inputs for RF-tend (which is analogous to NN1) are similar, 
except that in RF-tend all $48$ vertical levels are used as inputs. 
Furthermore, RF-tend predicts the net tendency summed over all processes included in NN1 of $H_{\rm{L}}$ and $q_{\rm{T}}$ 
at all $48$ vertical levels (with the exception that radiative heating was predicted up to $z=11.8$km), 
rather than predicting separate fluxes or tendencies for different physical processes.
As a test, we also tried to use NN1 with a similar input and output structure to RF-tend and thus have it predict the net tendency over all processes rather than separate fluxes and sinks for each process. However, this approach does not ensure physical consistency in an NN parameterization (see section~3.2 of the main paper) unlike an RF which predict averages over training samples and thus automatically conserves energy, and the quality of online results varied substantially when different grid spacings were used.  
Future work could further compare two identical structures of RFs and NNs.
Further details about RF-tend can be found in the supplementary information of YOG20.



%
%
%
%

\begin{figure}
\begin{centering}
\includegraphics[scale=1.1]{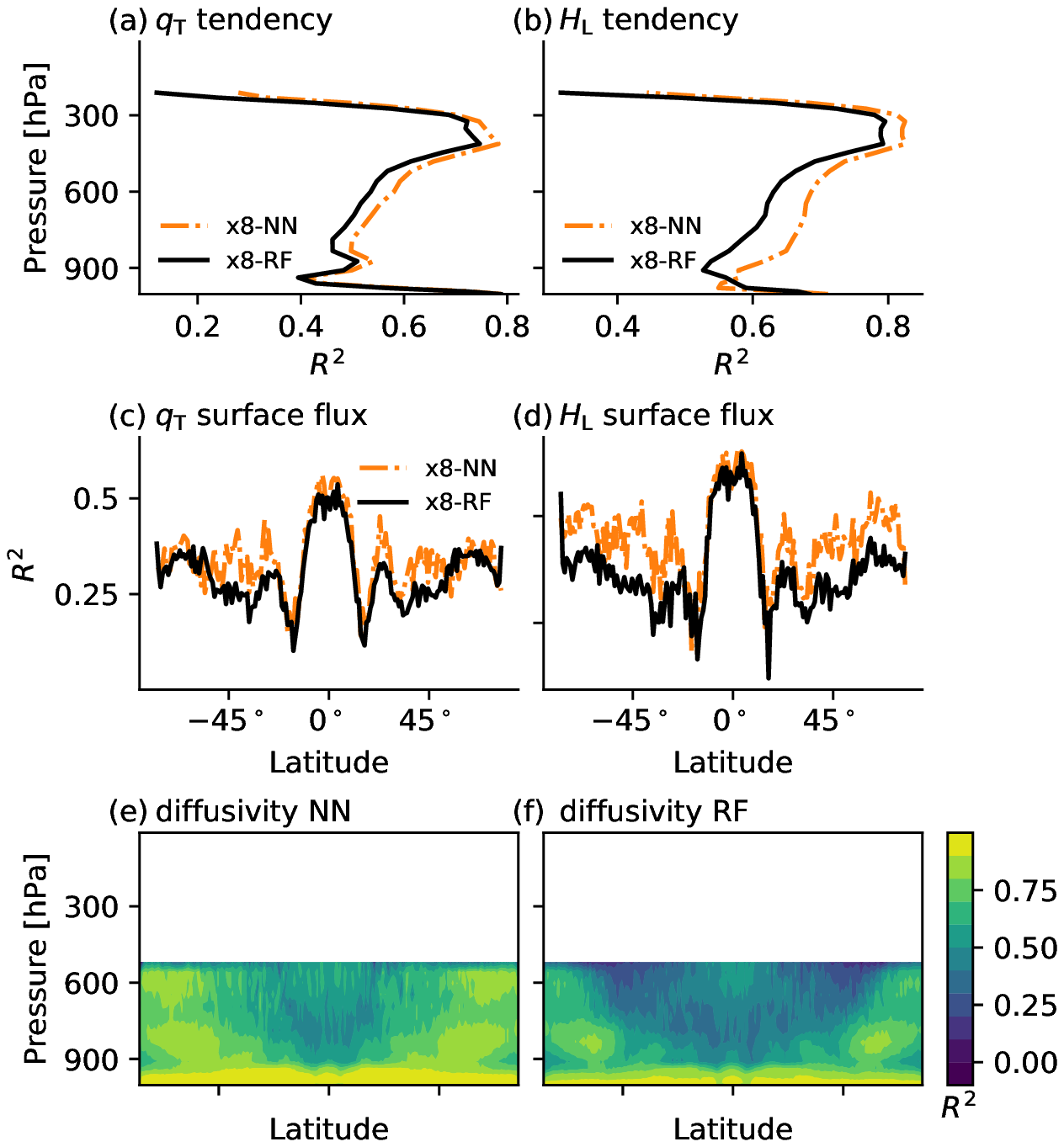} 
\par\end{centering}
\protect\caption{{ Coefficient of determination ($\rm{R}^2$) for offline performance of the x8-NN and x8-RF parameterizations for the 
(a) subgrid tendency of non-precipitating water mixing ratio ($q_{\rm{T}}$) as a function of pressure, 
(b) subgrid tendency of liquid/ice water static stability energy ($H_{\rm{L}}$) as function of pressure, 
(c) subgrid surface energy flux as a function of latitude,
(d) subgrid surface moisture flux as a function of latitude,
(e-f) diffusivity as a function of latitude and pressure for the (e) NN and (f) RF.
In panels a-d orange dash-dotted lines show results for NN and black lines show results for RF.
$\rm{R}^2$ is calculated based on the samples from the test datasets. }  \label{fig:Offline_Rsq}}
\end{figure}

NN2 and RF-diff have the same unscaled outputs and their offline results
 can be easily compared directly. 
 To compare the offline results between RF-tend and NN1 we  
use the unscaled tendencies of $H_{\rm{L}}$ and $q_{\rm{T}}$
as calculated from the outputs predicted by NN1, 
such that we compare the same target values. 
Offline performance is measured here 
as the coefficient of determination ($R^2$) as applied to the test dataset.
We find that the NN parameterization outperforms the RF parameterization 
across all variables (Figure~\ref{fig:Offline_Rsq}). 

 \textcolor{black}{The RF and NN parameterizations have some differences in their test (text~S2) and train datasets.
Since the RF parameterization requires more memory during training we 
used less training samples (5,000,000 for x8-RF compared to 13,856,040 for x8-NN).  Using the same number of samples as in the  NN lead to memory errors when training on nodes with 108Gb of random-access memory. Since the $R^2$ of the RF parameterization is roughly constant when increasing the number of training samples above 5,000,000 (Fig.~S8 in YOG20), it is unlikely that adding more samples would substantially change the offline result of RF parameterization.}


%
%


%
%
%
%

\bibliography{yanibib_long_names}

%


%
%
%
%
%

%
%
\clearpage


%
%
%
%

%
%
%
%
%

\begin{figure}
\centerline{\includegraphics{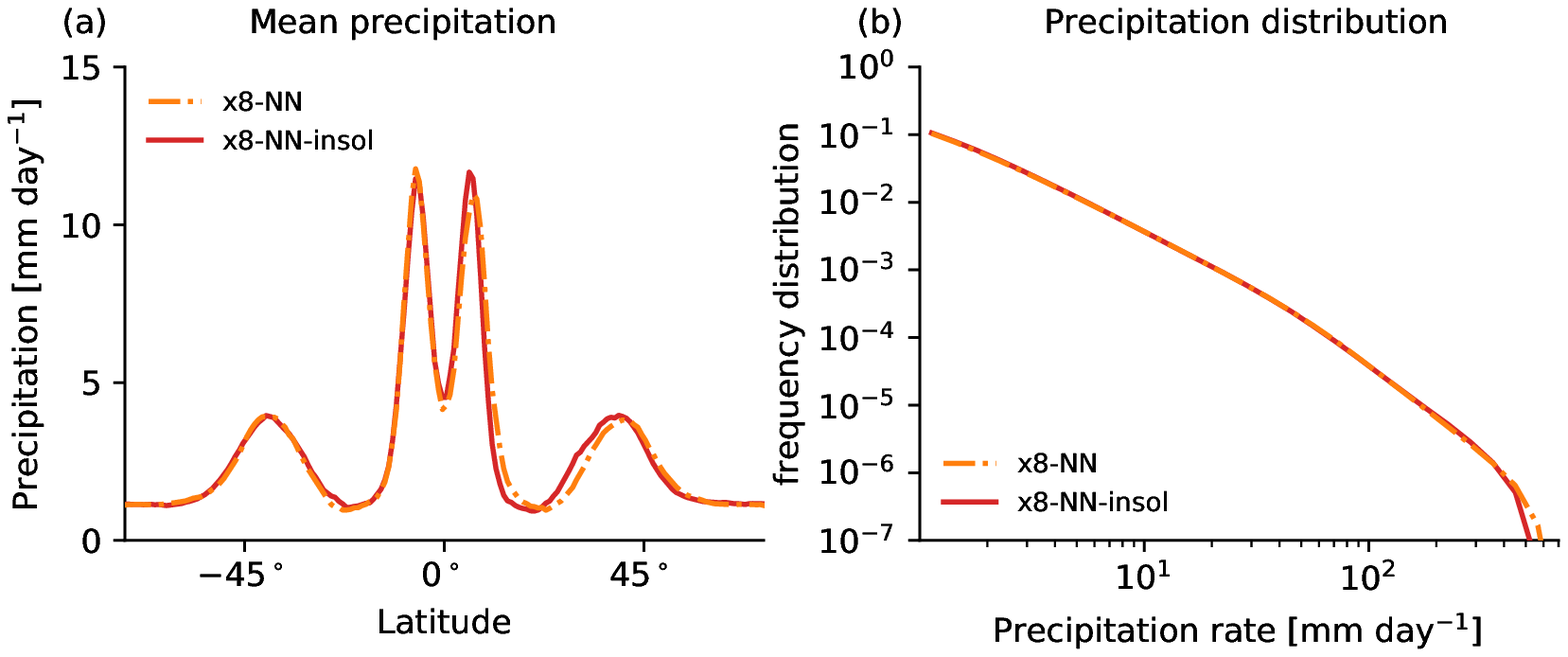}}
\caption{ 
(a) Zonal- and time-mean precipitation rate as a function of latitude and (b) frequency distribution of 3-hourly precipitation rate for coarse-resolution simulations with the default NN parameterization which uses distance to the equator as an input feature (x8-NN; orange dash-dotted) and an alternative NN parameterization that instead uses top of atmosphere insolation as an input feature (x8-NN-insol; red).}
\label{fig:precip_online_x8_solin}
\end{figure}

\begin{figure}
\centerline{\includegraphics{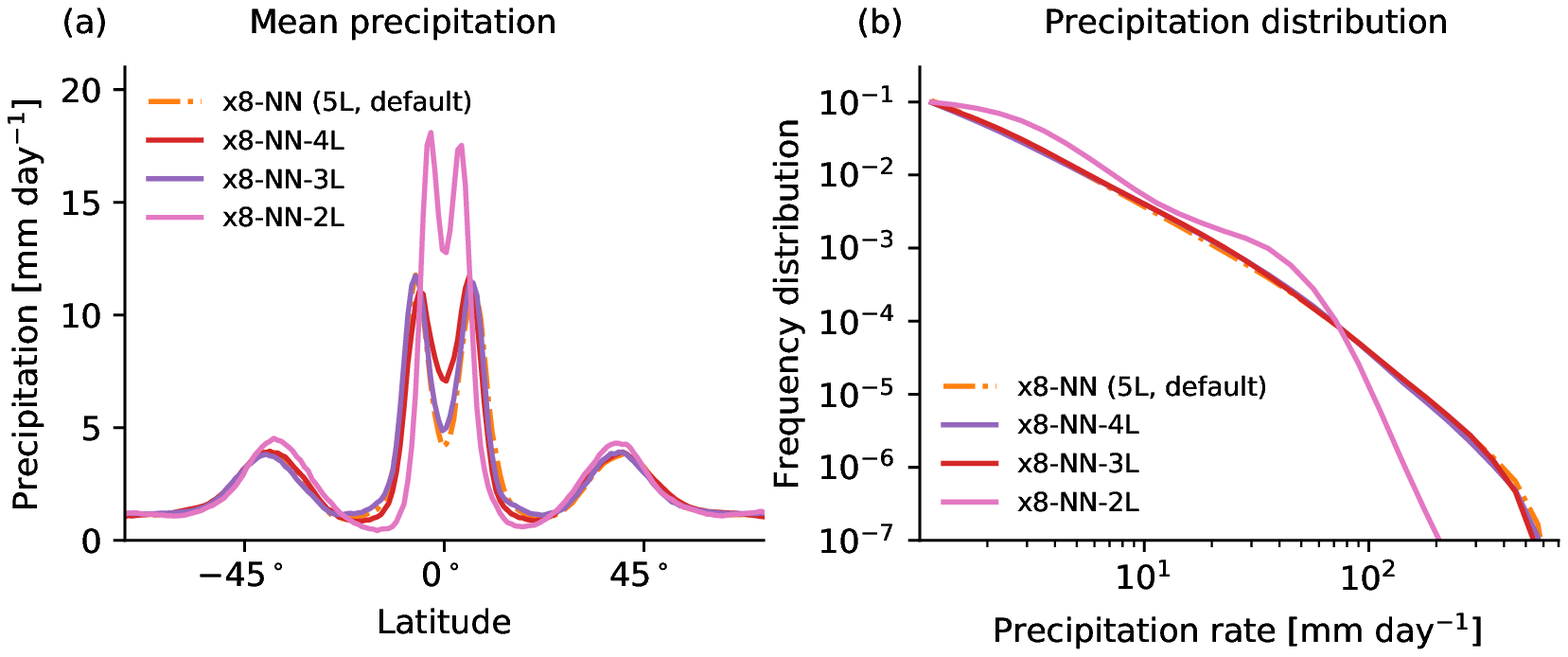}}
\caption{ 
(a) Zonal- and time-mean precipitation rate as a function of latitude and (b) frequency distribution of 3-hourly precipitation rate for the coarse-resolution simulation with the NN parameterization using 5 layers which is the default NN used in the paper (x8-NN; orange dash-dotted), 4 layers (x8-NN-4L; red), 3 layers (x8-NN-3L; purple) and 2 layers (x8-NN-2L; purple).}
\label{fig:precip_online_x8_layers}
\end{figure}

\begin{figure}
\centerline{\includegraphics{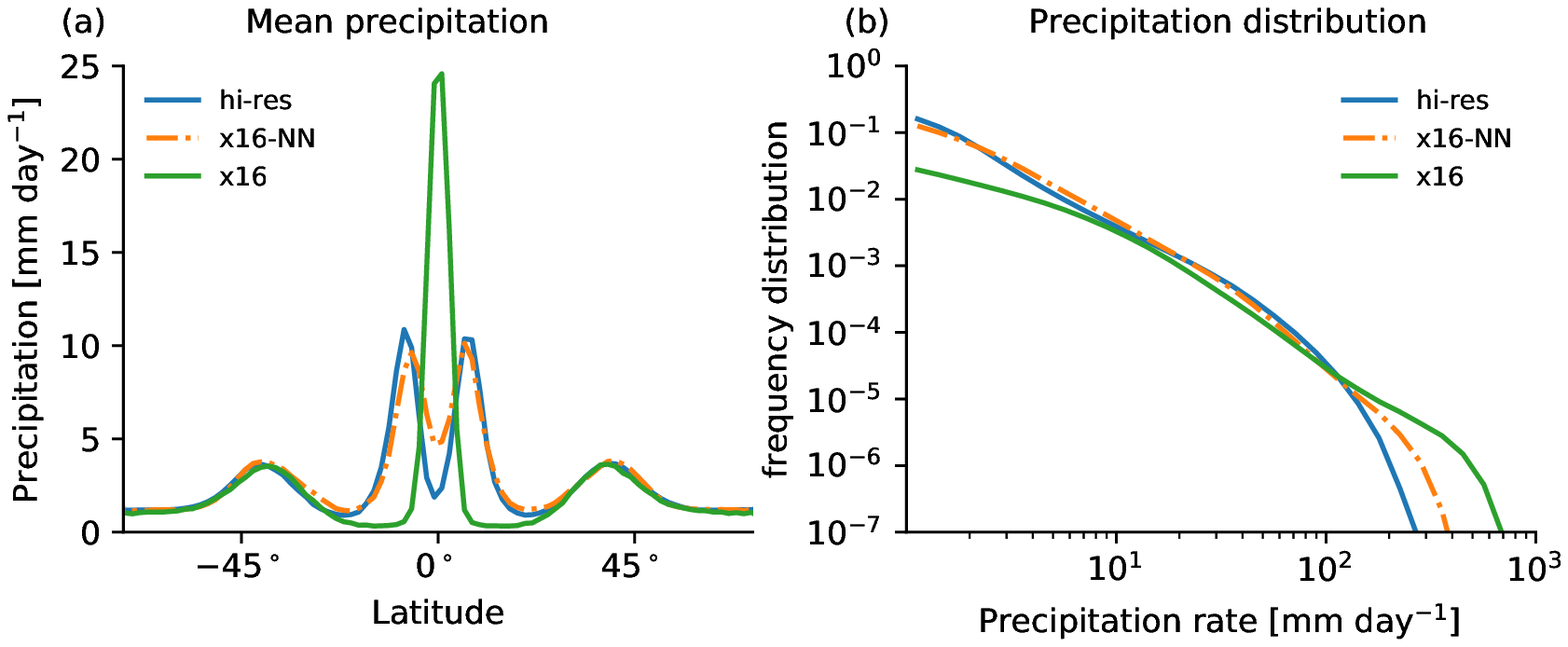}}
\caption{ 
(a) Zonal- and time-mean precipitation rate as a function of latitude and (b) frequency distribution of 3-hourly precipitation rate for the high-resolution simulation (hi-res; blue), the coarse-resolution simulation at 192km horizontal grid spacing without the NN parameterization (x16; green) and with the NN parameterization (x16-NN; orange dash-dotted).}
\label{fig:precip_online_x16}
\end{figure}

\begin{table}[]
\begin{tabular}{l|ccc|cccc} \hline      
 &   \multicolumn{3}{c|}{$R^2$ relative to hi-res}&  \multicolumn{4}{c}{$R^2$ relative to x8-NN}  \\ \hline
 &   x8 &  x8-RF & x8-NN  & x8-NN-7bits  &  x8-NN-5bits  & x8-NN-3bits  & x8-NN-1bits   \\ \hline
Eddy kinetic energy & 0.88  & 0.97                           & 0.94    & 0.99  & 0.99  & 0.99  & 0.30        \\
Zonal wind         & 0.87     & 0.98                           & 0.93       & 0.99& 0.99& 0.99  & 0.70 \\
Meridional wind   & -0.01  & 0.87                           & 0.80      & 0.98 & 0.99 & 0.95 & 0.78\\
Non-precipitating water & 0.97  & 0.99                 & 0.99    & 0.99 &  0.99&  0.99&  0.98 \\ 
Precipitation&-3.47  & 0.92                 & 0.88   &  0.94 & 0.99 & 0.86 & 0.79 \\ 
Precip. frequency dist. & 0.35  & 0.98                 & 0.99    & 0.99& 0.99& 0.99 &0.95 \\ \hline
\end{tabular}

\caption{
Online performance as measured by the coefficient of determination ($R^2$) of the time- and zonal-mean  of eddy kinetic energy, zonal wind, meridional wind, non-precipitating water, precipitation, and of the frequency distribution of 3-hourly precipitation. $R^2$ is calculated relative to hi-res for the coarse-resolution simulation with no ML parameterization (x8), with the RF parameterization (x8-RF) and with the NN parameterization (x8-NN). $R^2$ is calculated relative to x8-NN for the simulations with reduced-precision parameterizations with 7 (x8-NN-7bits), 5 (x8-NN-5bits), 3 (x8-NN-3bits) and 1 (x8-NN-1bits) bits in the mantissa. 
The eddy kinetic energy is defined with respect to the zonal and time mean.
x8-NN-7bits corresponds to the bfloat16 half-precision format which has 7 bits in the mantissa, and the default parameterization (x8-NN) is single precision which has 23 bits in the mantissa.
}
\label{table:Rsq_yz_online_eke_u_v_qv}
\end{table}

\begin{table}[]
\centering
\begin{tabular}{lccccc}
\hline
       & $(H_{\rm{L}})_{\rm{adv}}^{\rm{subg-flux}}$ & $(q_{\rm{T}})_{\rm{adv}}^{\rm{subg-flux}}$ & $(q_{\rm{T}})_{\rm{sed}}^{\rm{subg-flux}}$ & $(q_{\rm{T}})_{\rm{mic}}^{\rm{tend}}$ & $(H_{\rm{L}})_{\rm{rad-phase}}^{\rm{tend}}$ \\ \hline
x8-NN-2L                      & 0.59              & 0.60                    & 0.67  & 0.68    & 0.59              \\
x8-NN-3L                      & 0.62              & 0.64                    & 0.77  & 0.77    & 0.73             \\
x8-NN-4L                      & 0.62              & 0.65                    & 0.79  & 0.78    & 0.76              \\
x8-NN (5L)                    & 0.63               & 0.65                    & 0.80    & 0.79      & 0.76        \\ 
x16-NN (5L)                     & 0.69              & 0.75                    & 0.84    & 0.80      & 0.79        \\ \hline
\end{tabular}
\caption{ 
Offline performance of NN1 for different NN architectures and for coarse-graining factor as measured by $R^2$. The offline performance is given for different outputs for  NNs with 2,3,4 and 5 layers at x8 resolution (x8-NN-2L, x8-NN-3L, x8-NN-4L, x8-NN, respectively), and for x16 resolution with 5 layers (x16-NN). Note that x8-NN and x16-NN use the default of 5 layers.
All vertical levels used are included when calculating $R^2$. All results are based on the test dataset.}
\label{table:offline performance}
\end{table}